# A Diamagnetic, Light-Driven Tesla Engine Based on a Mechanically Displaced, Magnetically Levitated Graphene Disk

Tian Tong[1,2†], Feng Lin[2,3†], Wei Zhang[4], Runjia Li[5], Xinxin Xing[2,3], Zhuochen Duan[3], Chunhui Xu[3], Bing Tu[1], Zhaoping Liu[6], Xufeng Zhou[6], Zhiming Wang[7], Dong Liu[5], Jonathan Hu[4] and Jiming Bao[2]*

[1]School of Physics and Optoelectronic Engineering, State Key Laboratory Cultivation Base of Atmospheric Optoelectronic Detection and Information Fusion, Nanjing University of Information Science and Technology, Nanjing, Jiangsu 210044, China

[2]Department of Electrical & Computer Engineering and Texas Center for Superconductivity (TcSUH), University of Houston, Houston, Texas 77204, USA

[3]School of Materials and Energy, Yunnan University, Kunming, Yunnan 650091, China

[4]Department of Electrical and Computer Engineering, Baylor University, Waco, Texas 76798, USA

[5]Department of Mechanical Engineering, University of Houston, Houston, Texas 77204, USA

[6]Ningbo Institute of Materials Technology and Engineering, Chinese Academy of Sciences, Ningbo, Zhejiang 315201, China

[7]Shimmer Center, Tianfu Jiangxi Laboratory, Chengdu, Sichuan 641419, China

[†]T. Tong and F. Lin contributed equally to this work.

*To whom correspondence should be addressed. Emails: jbao@uh.edu.

# Abstract

Ferromagnetic materials are widely used in Tesla thermomagnetic engines, whereas diamagnetic counterparts have remained unexplored. Here, we demonstrate the first diamagnetic Tesla engine by exploiting the strong diamagnetism of graphene. A graphene disk, fabricated by stacking graphene sheets, serves as the engine wheel. We first show that the conventional Tesla engine design using a permanent magnet placed near the disk edge to create unbalanced thermomagnetic forces under asymmetric local heating fails to generate rotation. We achieve stable operation by laterally displacing the levitated disk from equilibrium, creating a strong restoring force that drives rotation under light excitation. Calculations and measurements establish the displacement-dependent force, with an optimal offset of 0.8 mm yielding speeds up to 2000 rpm under laser heating and 1000 rpm under direct sunlight. Adding vanes allows the disk to function as a gear, powering a graphene vehicle and transferring energy to another disk. This design utilizes the strong and anisotropic diamagnetism of graphene and paves the way for light-powered sensors, actuators, and micro-vehicles.

**Introduction**

Tesla thermomagnetic engines are heat engines capable of converting thermal energy into continuous mechanical work through a magnetically driven wheel (*1-5*). Prior to the recent demonstration of a liquid thermomagnetic engine based on a paramagnetic material (*6*), conventional solid-state thermomagnetic engines employed wheels or disks made of ferromagnetic materials positioned adjacent to a permanent magnet, generating a strong magnetic field parallel to the disk plane. In the absence of heating, the entire disk maintains uniform magnetic susceptibility, producing balanced magnetic attraction forces from the two half-regions on either side of the line connecting the centers of the disk and the magnet. When one region of the disk is heated, its magnetic susceptibility decreases while the susceptibility of the unheated other region remains unchanged, resulting in an unbalanced magnetic force that drives rotation of the wheel (*1-5*). Because the magnetic susceptibility of all magnetic materials changes with temperature, the operating principle of thermomagnetic engines is, in principle, also applicable to diamagnetic materials. However, Tesla thermomagnetic engines based on diamagnetic materials have not been reported to date.

In this work, we demonstrate the first Tesla thermomagnetic engine based on a diamagnetic material. Graphene was selected because of its strong and anisotropic diamagnetic susceptibility (*7, 8*), Single- and few-layer graphene sheets were magnetically aligned and stacked to form a graphene disk serving as the engine wheel (*9-11*). However, attempts to operate the device using the conventional configuration—with a permanent magnet placed parallel to the graphene disk—were unsuccessful. Inspired by our recent demonstration of a ferrofluid Tesla engine (*6*), we developed a new design in which the magnetic field is oriented perpendicular to the graphene disk. In this configuration, the graphene disk is magnetically levitated and laterally displaced, a geometry that both eliminates rotational friction and maximizes the thermomagnetic driving force. Through combined theoretical modeling and experimental characterization, we quantified the displacement-dependent diamagnetic confinement force and identified an optimal lateral offset for stable, high-speed operation. Using this design, we achieved rotation speeds of 2000 rpm under laser heating and 1000 rpm under direct sunlight. Furthermore, we demonstrated the versatility of this graphene Tesla engine by propelling a levitated graphene foil along a straight magnetic track and by driving a graphene gear constructed from a disk with attached vanes.

**Results and Discussion**

We selected graphene for the Tesla thermomagnetic engine because it possesses a combination of favorable and unique properties, including large diamagnetic susceptibility, strong optical absorption, exceptional mechanical strength, and high thermal conductivity (*12*). Graphene is well known for its high in-plane electrical conductivity and electron mobility, which give rise to its unusually strong out-of-plane diamagnetic response—second only to superconductors (*7, 8*). This strong but anisotropic diamagnetic susceptibility enables magnetic alignment and orientation control of graphene sheets (*9-11*), in addition to magnetic levitation (*13*). To retain these intrinsic properties while producing a practical wheel for the Tesla engine, we used both liquid exfoliation and electrochemical exfoliation to prepare single- and few-layer graphene sheets, which were subsequently magnetically aligned and stacked into graphene foils (*13, 14*). These two-dimensional foils preserve the essential characteristics of graphene, while allowing precise control over size and geometry. In principle, graphene disks of arbitrary diameter and shape can be fabricated using this approach (*13, 14*).

To demonstrate a graphene-based Tesla thermomagnetic engine, our initial approach followed the conventional design of ferromagnetic-material Tesla engines by replacing the ferromagnetic wheel with a graphene disk (*1-5*). As shown in Fig. 1a, the graphene disk is suspended by a horizontal copper wire passing through its central hole, with its edge aligned perpendicular to the surface of a cubic magnet. Unlike ferromagnetic or paramagnetic materials, graphene—as a diamagnetic material—experiences a repulsive force rather than an attractive force in a magnetic field, as shown in Fig. 1b. This repulsive force was observed when we moved the magnet close to the edge of the disk. Theoretically, when a laser beam illuminates the graphene disk asymmetrically with respect to the yellow dashed line connecting the centers of the graphene disk and the magnet (Fig. 1a–c), the local increase in temperature reduces the magnetic susceptibility in the illuminated region. This reduction in susceptibility decreases the local repulsive force and produces an anticlockwise torque (*7, 8, 13*). However, despite the low mass of the graphene disk, no rotational motion was observed. This is likely due to the weak thermomagnetic force, deviations from ideal disk circularity, and possible friction between the copper wire and the disk's central hole.

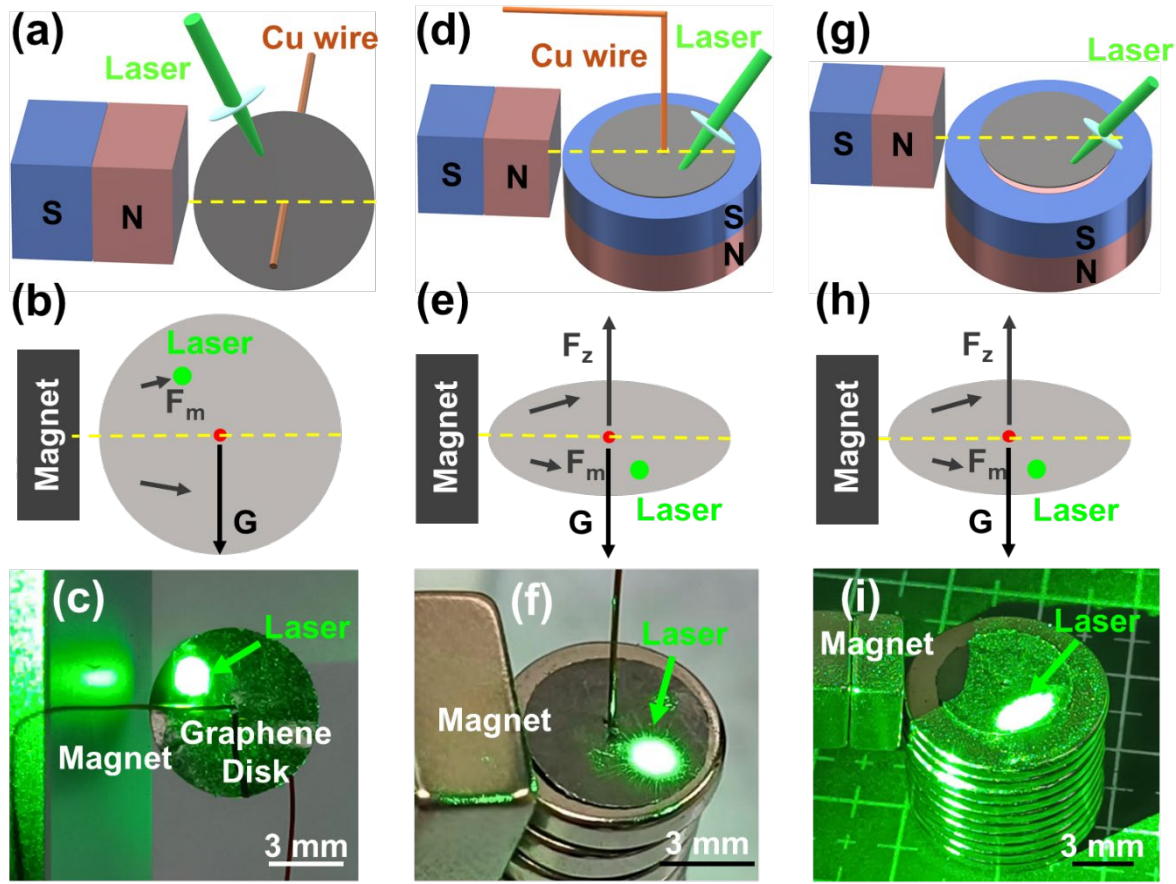

**Fig. 1.** Attempts and failures of Tesla engine designs using traditional configurations. (a-c) Conventional design in which a cubic magnet is placed near the edge of a graphene disk. (d-f) Revised design using a cylindrical magnet to levitate the graphene disk, with a copper wire used to position the disk at the magnet's center. (g-i) Further revised design employing a cylindrical–ring magnet combination to achieve more stable levitation of the graphene disk. The 532 nm laser power is 1 W.

To eliminate possible mechanical resistance, we used a cylindrical magnet to levitate the graphene disk by exploiting its strong out-of-plane diamagnetic response. Fig. 1d–e show this revised design. A copper wire is used to hold the disk at the center of the magnet to prevent it from slipping sideways (*15-17*). In principle, this configuration appears promising: the cylindrical magnet provides levitation, while the bar magnet supplies the rotational torque. However, in practice, no rotation of the graphene disk is observed. One contributing factor is the interaction between the two magnets, which destabilizes the levitation. As shown in Fig. 1f, the disk becomes tilted and makes contact with the cylindrical magnet. To improve stability, we further modified the design by employing concentric cylindrical–ring magnets, which provide more robust levitation of the graphene disk (*16, 17*), as shown in Fig. 1g–h. Nevertheless, as shown in Fig. 1i, the disk remains slightly tilted due to interaction with the two magnets, and no rotation is observed.

The failure of the traditional Tesla engine designs for a diamagnetic disk prompted us to re-examine the underlying operating principles. In all three configurations shown in Fig. 1, a magnet positioned at the disk edge appears essential, as this arrangement is standard in Tesla engines based on ferromagnetic materials (*1-5*). In that context, the edge magnet provides the unbalanced thermomagnetic force that drives rotation. However, this is not the only configuration capable of producing such an unbalanced force. In our previously developed ferrofluid Tesla engine, the rotational force is not generated by an edge magnet. Instead, a cylindrical magnet placed beneath the ferrofluid provides a magnetic field perpendicular to the fluid surface (*6*). When the ferrofluid is displaced from the field center, a lateral magnetic force arises, producing the torque needed for rotation. Notice that the concentric cylindrical–ring magnet combination used in Fig. 1g provides stable levitation of the graphene disk (*16, 17*). A slight displacement of the disk from the magnetic center naturally induces a restoring lateral force within the graphene plane. Thus, a magnet placed near the disk edge is not necessary to generate the driving torque. Guided by this reasoning, we developed a new design. As shown in Fig. 2a, we removed the edge magnet and instead used the concentric cylindrical–ring magnet pair solely for levitation. We then intentionally introduced a thin copper wire—passing through the disk's central hole—to slightly displace the disk from the magnetic center, thereby generating a restoring magnetic force that can be exploited to drive rotation.

To quantify this force and find out the optimal displacement for the Tesla engine, we calculated the in-plane magnetic force on the graphene disk and its dependence on the magnetic field's intensity and direction. The effective induced magnetization **M** with the magnetic susceptibility tensor $\chi$ in the presence of a static magnetic flux density **B** can be expressed as (*13, 15, 16, 18*)

$$\mathbf{M} = \frac{1}{\mu_0}\chi \cdot \mathbf{B}, \qquad (1)$$

where $\mu_0$ is the vacuum permeability, and the force density $f$ can be expressed as

$$\mathbf{f} = (\mathbf{M}\cdot\nabla)\mathbf{B} = \frac{1}{2\mu_0}\nabla\left(\chi_x B_x^2 + \chi_y B_y^2 + \chi_z B_z^2\right), \qquad (2)$$

where $\chi_x$ and $\chi_y$ are equal to transverse magnetic susceptibility $\chi_\parallel = -0.85 \times 10^{-4}$ and $\chi_z$ is equal to perpendicular magnetic susceptibility $\chi_\perp = -4.5 \times 10^{-4}$. Adopting an approximation approach similar to that in our previous work (*13*), we first determine the levitation height of the graphene disk without displacement, then shift the disk horizontally to calculate the in-plane force based on the magnetic flux density distribution at this levitation height. Our approach differs from that of Huang et al., which assesses stability by dividing the disk into magnetic regions and calculate the resulting magnetic potential energy (*16*). Typically, the graphene disk has a diameter of 6 mm and a mass of 1.7 mg, the cylindrical magnets—each with a diameter of 5 mm—are assembled inside the concentric ring magnets (which have an inner diameter of 5 mm and an outer diameter of 10 mm), and the total magnetic flux density at the center at z = 0.1 mm is 600 mT. As shown in Fig. 2b, the dotted line indicates the gravity of the graphene disk, intersecting with the curve of magnetic force at *z*=0.474 mm, which represents the levitation height of the graphene disk. We further calculated the magnetic flux density in x, y, and z directions in *z*=0.474 mm plane, the simulation model and the orientation of the coordinate axes are illustrated in Fig. 2c. The distributions of magnetic flux density ($B_x$, $B_y$, $B_z$) are separately illustrated in Fig. 2d, 2e, and 2f, respectively.

Based on the magnetic-field distributions, we calculated the variation of the magnetic force on the graphene disk as a function of lateral displacement along the x-axis at a levitation height of 0.474 mm. As shown in Fig. 2g, the magnetic force increases from 0 to $-1.5 \times 10^{-6}$ N as the disk shifts from 0 to 0.8 mm, after which the force begins to decrease and eventually reverses direction. The operating mechanism of this design can therefore be understood as follows. As illustrated in Fig. 2h, when the disk is displaced from the magnetic center, the net magnetic force acting on it is a restoring force F that pulls it back toward the equilibrium position—the center of the magnetic field where the forces are balanced. For analysis, we divide the graphene disk symmetrically into halves A and B. The restoring forces on each half (**F**$_A$ and **F**$_B$), shown in Fig. 2i, are equal in magnitude and their vector sum yields the total restoring force **F**. When the laser irradiates part B (Fig. 2j), the resulting temperature rise ($T_B > T_A$) reduces the susceptibility of graphene. This weakens both **F**$_A$ and **F**$_B$, but **F**$_B$ decreases more due to higher heating. The imbalance (|**F**$_A$| > |**F**$_B$|) creates a net magnetic torque that drives clockwise rotation. Furthermore, we also investigated the contribution of magnetic flux density in different directions to the in-plane magnetic force acting on the graphene disk. Fig. 2k-n show the simulation results of horizontal magnetic force density $f_x$ due to magnetic fields at different orientations. Comparison among the figures indicates that the z-component of magnetic flux density $B_z$ is the primary contributor to the horizontal magnetic force density $f_x$. This finding differs from intuitive expectation and was not recognized in previous calculations (*16*), proving that the in-plane (horizontal) magnetic force acting on the graphene disk originates from the out-of-plane (vertical) magnetic field component. It also provides a compelling explanation for the lack of rotational motion in the first two designs presented in Fig. 1a and 1d, and highlights the ingenuity embedded in our optimized design—the vertical component of the magnetic field simultaneously levitates the graphene disk and generates the driving force for its rotation.

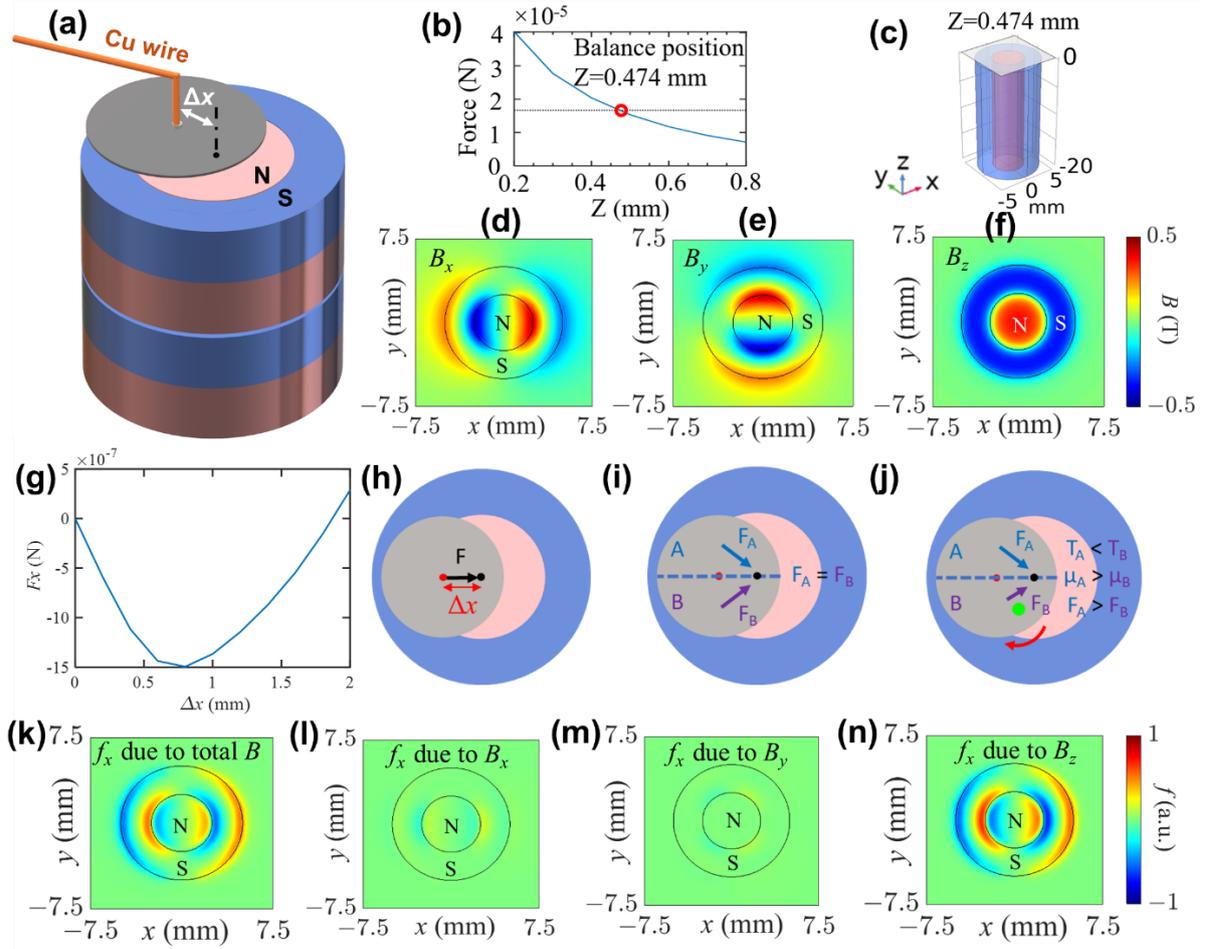

**Fig. 2.** New design and calculation of restoring force as a function of lateral displacement. (a) Schematic of the new graphene Tesla thermomagnetic engine. A magnetically levitated graphene disk is laterally displaced by a copper wire passing through its central hole. (b) Calculated vertical magnetic force on the graphene disk (6 mm diameter, 1.7 mg) at different heights along the z-axis (magnetic field: 600 mT at z = 0.1 mm). The dotted line indicates the gravitational force on the disk, assuming no lateral displacement. (c) Schematic of the magnetic-field simulation model on the plane 0.474 mm above the magnets. (d–f) Normalized magnetic-field distributions in the x, y, and z directions at z = 0.474 mm. (g) Simulated horizontal magnetic force on the graphene disk as a function of lateral displacement along the x-axis, assuming a levitation height of z = 0.474 mm. (h–j) Schematics illustrating the rotation principle and force analysis. (k–n) Normalized contributions to the horizontal magnetic-force density $f_x$ from magnetic-field components in different directions: (k) total field, (l) x-component, (m) y-component, and (n) z-component.

We further performed experimental measurements of the horizontal magnetic force to validate the simulation results. Fig. 3a–c show the experimental setup. A long optical fiber was fixed to a translation stage, and its free end was passed through the central hole of the graphene disk. The bending of the fiber was used to measure the restoring force; for small bending angles, the force is proportional to the deflection angle (*19*). Fig. 3d–e present the measured restoring magnetic force as a function of lateral displacement. The force increases nearly linearly at small displacements, then grows sub-linearly and reaches a maximum at approximately 0.8 mm, after which it decreases. This experimental trend agrees well with the simulations shown in Fig. 2. Although the absolute magnitude of the force was not calibrated, this does not affect the validity of the relative force measurements or the displacement dependence.

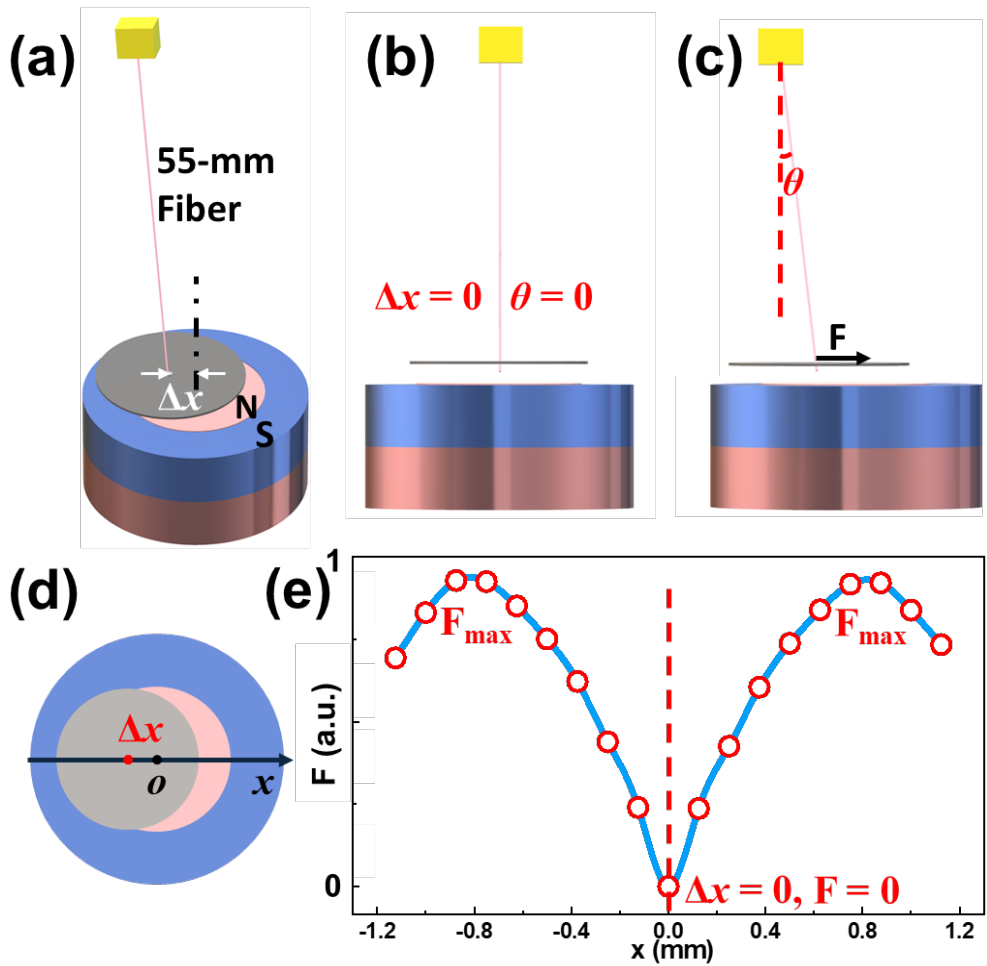

**Fig. 3.** Experimental measurement of magnetic force as a function of lateral displacement. (a) Experimental setup for measuring small magnetic forces. (b–c) Side views of the setup: (b) initial equilibrium position and (c) position after a lateral displacement $\Delta x$. (d) Top view of the experimental setup, where O denotes the center of the magnetic field. (e) Measured magnetic force F as a function of lateral displacement.

Supported by both the calculated and experimentally measured magnetic restoring forces, we conducted a new set of experiments. A copper wire was used to displace the graphene disk slightly from the magnetic center of the cylindrical–ring magnet pair, after which the disk was illuminated. As shown in Fig. 4a–d and supporting videos SV1-2, the new Tesla engine begins to rotate as soon as light from either a laser or sunlight is introduced. The rotation direction is precisely controlled by the combination of disk displacement and the location of light illumination. For example, in Fig. 4a–b, when the disk is pulled to the right and the laser illuminates the lower portion, the temperature of the lower half increases, reducing the restoring force on that side. Meanwhile, the restoring force on the upper half remains unchanged, generating an anticlockwise torque and resulting in rotation. Only snapshots are shown here; rotation speeds were extracted from these images and the corresponding videos. The uneven temperature distribution responsible for the torque can be directly verified using an IR camera (*6, 20*). As shown in Fig. 4e–j, the laser spot and its surrounding region exhibit a higher temperature than the rest of the disk, even when rotating at high speed. This continually evolving but asymmetric temperature profile sustains the rotational motion.

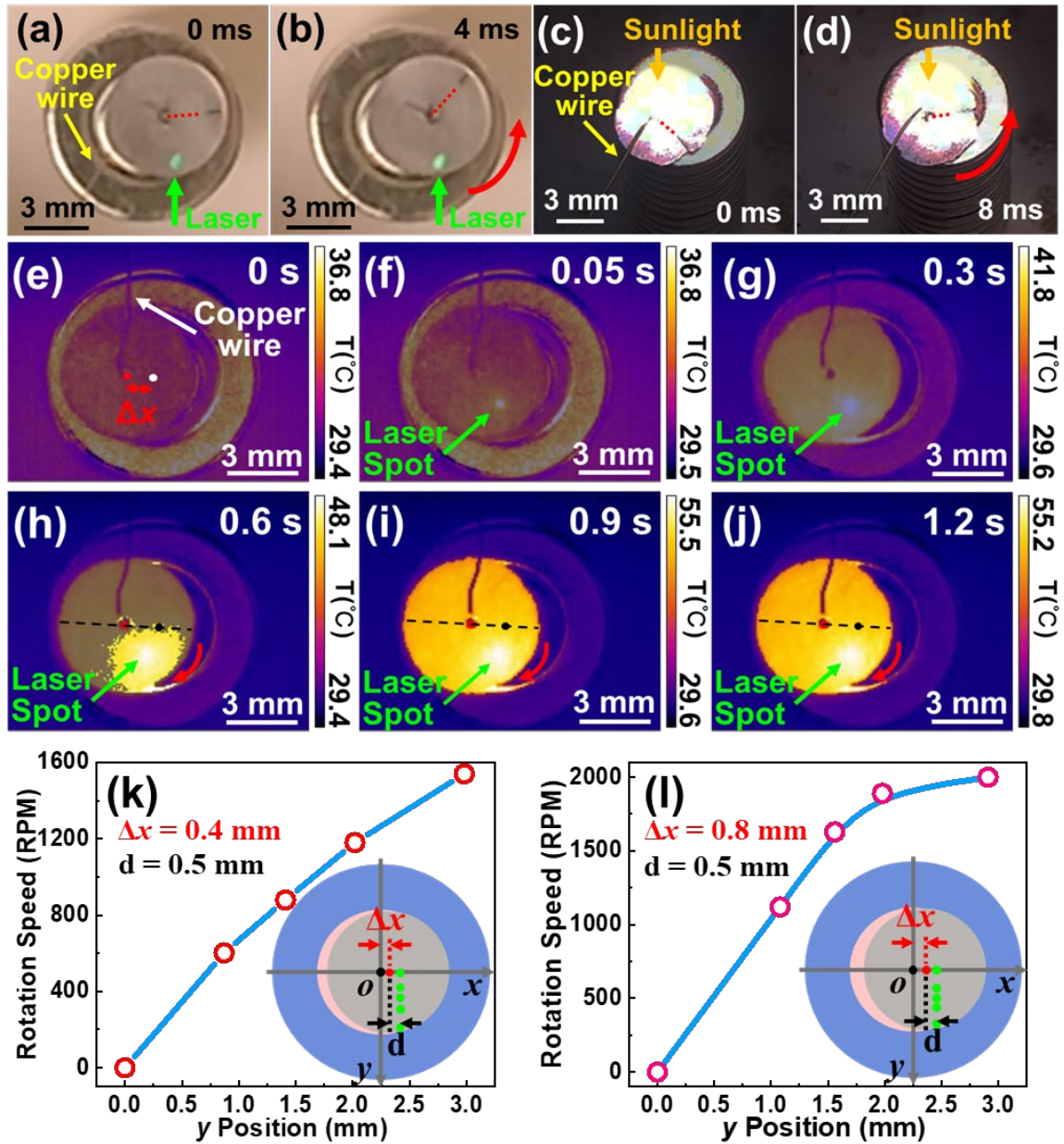

**Fig. 4.** Demonstrations of the graphene Tesla thermomagnetic engine. (a-b) Two consecutive snapshots of disk rotation under laser illumination. The red dashed line marks a reference point on the disk. (c-d) Two consecutive snapshots of disk rotation under sunlight illumination, with the red dashed line indicating the reference mark. (e-j) Infrared thermal images of the graphene engine under laser illumination at (e) 0 s, (f) 0.05 s, (g) 0.3 s, (h) 0.6 s, (i) 0.9 s, and (j) 1.2 s. (k-l) Rotation speed as a function of the y-displacement of the laser spot for x-displacements of the graphene disk of (k) 0.4 mm and (l) 0.8 mm. Inset: relative position of the laser spot (100 μm diameter). The x-displacement between the disk center and the laser spot was kept at 0.5 mm.

To investigate how the rotation speed depends on the disk displacement and the laser irradiation

position—and to determine the conditions for achieving maximum speed—we fixed the laser power at 200 mW and varied the irradiation position on the graphene disk, as shown in the insets of Fig. 4k and 4l. When the laser spot was aligned along the x-axis, no rotation was observed because the heating was symmetric. We then selected two x-displacements: 0.8 mm, which corresponds to the maximum restoring force, and 0.4 mm, which is half of that value. For each displacement, the laser spot was scanned along the y-axis. Fig. 4k and 4l show that the rotation speed increases in both cases as the laser spot moves away from the x-axis, with the highest speed occurring at the 0.8 mm x-displacement. At the maximum y-displacement, the rotation speed reaches approximately 2000 rpm (Supporting video SV3). This behavior is expected: the 0.8 mm x-displacement provides the strongest restoring force, while the maximum y-offset produces the heating asymmetry at the maximum lever arm length. Together, the large restoring force and the maximal lever arm generate the highest rotation torque, resulting in the maximum rotation speed.

The demonstration of a graphene light-driven Tesla thermomagnetic engine opens exciting opportunities for device applications. Fig. 5a illustrates one such example, where rotational motion is converted into translational motion. We designed a graphene gear wheel by attaching two graphene vanes to a graphene disk and placed a rectangular graphene plate adjacent to it. The plate was levitated above three NdFeB bar magnets, allowing it to move freely along the magnetic track without mechanical friction. When the laser beam illuminated the graphene gear wheel, the wheel began to rotate and pushed the graphene plate horizontally along the track to the end of the magnets, as shown in Fig. 5a and supporting video SV4. The rotational motion can also be transmitted to another gear system. As shown in Fig. 5b and supporting video SV5, we positioned a four-vane graphene wheel next to the original two-vane wheel. Upon laser illumination of the first wheel, it began to rotate and transferred its mechanical energy to drive the second wheel. It is worth mentioning that all components in Fig. 5 are made entirely of graphene, whose high mechanical strength, ultralight weight, and strong diamagnetism enable magnetic levitation. These characteristics open intriguing possibilities for future applications across a variety of fields.

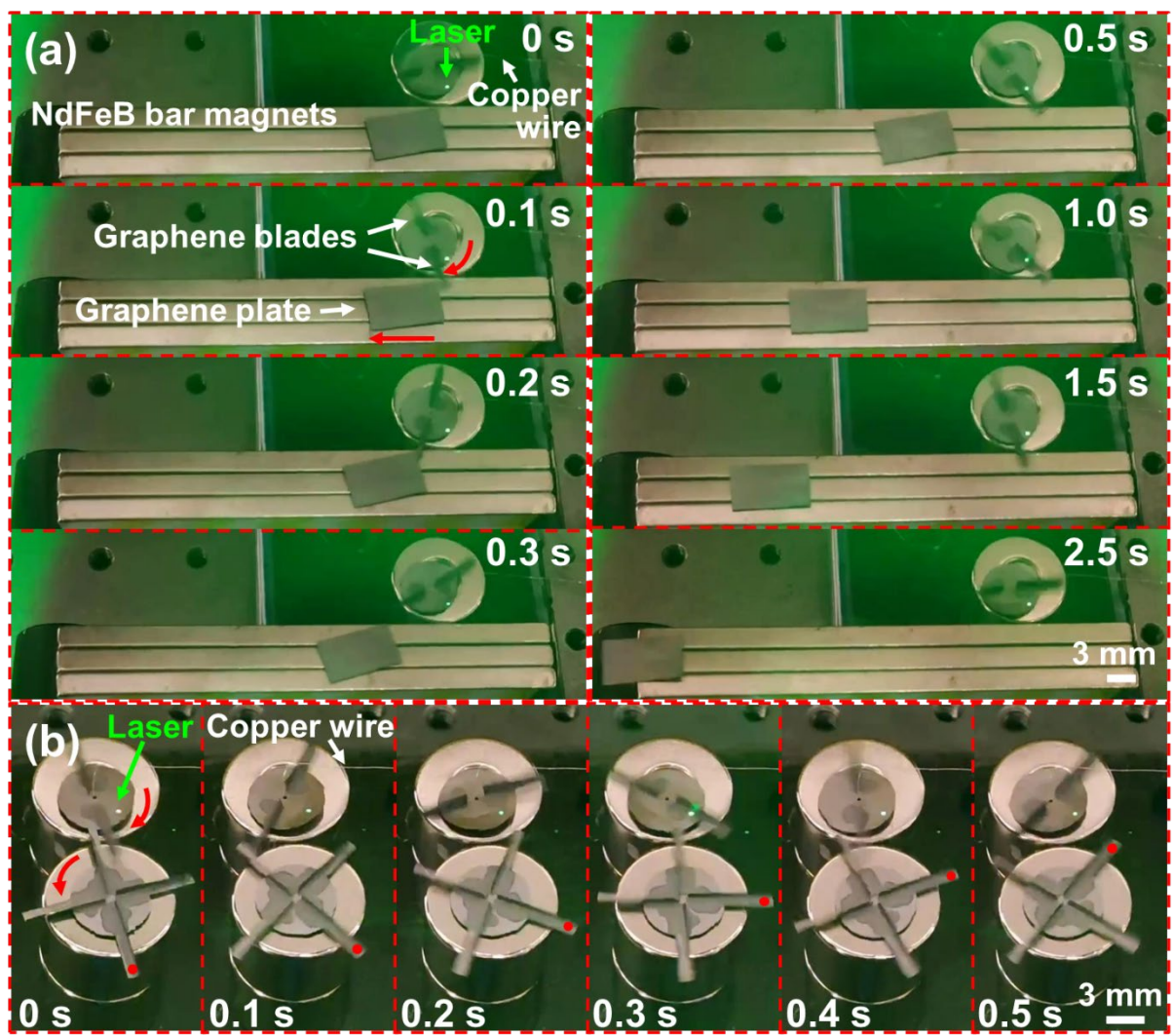

**Fig. 5.** Driving other graphene devices by the graphene Tesla engine. Snapshots showing mechanical motion transfer from the graphene Tesla engine to (a) a graphene vehicle levitated on a magnetic track and (b) a secondary graphene gear wheel.

**Conclusion**

In conclusion, we fabricated diamagnetic graphene disks from magnetically aligned graphene sheets and demonstrated the first diamagnetic Tesla engine using a graphene disk as the rotating wheel. We showed that the traditional Tesla engine design—placing a magnet at the edge of the wheel—fails for diamagnetic materials. Instead, our graphene Tesla engine employs a cylindrical–ring magnet combination that simultaneously provides stable magnetic levitation and thermomagnetic driving. For the first time, we identify that the perpendicular (out-of-plane) magnetic field component is responsible for both levitation and thermomagnetic actuation when the disk is displaced from the magnetic center. We calculated and experimentally measured the magnetic restoring force as a function of displacement, obtaining excellent agreement between theory and experiment. Under optimal displacement and laser heating conditions, the engine reached a maximum rotation speed of 2000 rpm. We further demonstrated that the graphene Tesla engine can drive other fully levitated graphene devices, including a graphene plate on a linear magnetic track and a secondary graphene vane above a cylindrical magnet. These demonstrations highlight the advantages of graphene—high mechanical strength, ultralow mass, and strong anisotropic diamagnetism—for frictionless, contact-free motion. Overall, the graphene Tesla engine establishes a new platform for light-driven rotational actuation in diamagnetic systems. This design sets a benchmark for future Tesla engines based on diamagnetic materials and opens new avenues for applications in light sensing, optomechanical actuators, temperature sensing, and micro-scale delivery vehicles, ultimately paving the way for a new class of light-powered graphene machines.

**Competing interest**

The authors declare that they have no known competing financial interests or personal relationships that could have appeared to influence the work reported in this paper.

**Data and materials availability**

The original data in this work is available from the corresponding authors upon reasonable requests.